\documentclass[preprint,byrevtex, prb,showpacs]{revtex4}
\usepackage{epsfig}
\usepackage{amsmath}
\usepackage{amssymb}
\usepackage{times}
\begin{document}
\title{Giant enhanced optical nonlinearity of colloidal nanocrystals with a graded-index host}
%\author{authors}
%\affiliation{Department of Physics, The Chinese University of Hong Kong, Shatin, New Territories, Hong Kong,
%China}
\author{J. J. Xiao\footnote{Electronic address: jjxiao@phy.cuhk.edu.hk}}
\affiliation{Department of Physics, The Chinese University of Hong Kong,
Shatin, New Territories, Hong Kong, China}
\author{K. W. Yu\footnote{Electronic address: kwyu@phy.cuhk.edu.hk}}
 \affiliation {Department of Physics, The Chinese University
of Hong Kong, Shatin, New Territories, Hong Kong, China}
\affiliation{Institute of Theoretical Physics, The Chinese University of Hong
Kong, Shatin, New Territories, Hong Kong, China}

\begin{abstract}
The effective linear and third-order nonlinear optical properties of metallic
colloidal crystal immersed in a graded-index host fluid are investigated
theoretically. The local electric fields are extracted self-consistently based
on the layer-to-layer interactions, which are readily given by the Lekner
summation method. The resultant optical absorption and nonlinearity
enhancement show a series of sharp peaks, which merge in a broadened resonant
band. The sharp peaks become a continuous band for increasing packing density
and number of layers. We believe that the sharp peaks arise from the in-plane
dipolar interactions and the surface plasmon resonance, whereas the continuous
band is due to the presence of the gradient in the host refractive index.
These results have not been observed in homogeneous and randomly-dispersed
colloids, and thus would be of great interest in optical nanomaterial
engineering.
\end{abstract}
\date{\today}
 \maketitle

There have been considerable interests in searching for materials with a large
nonlinear susceptibility. The most common way to achieve this is to use
composite materials in which the constituent components possess  large
intrinsic nonlinear responses. Noble metal (typically gold, silver and copper)
is often chosen as an ingredient due to their extremely large and fast
nonlinear optical response.
%as compared to other kind of organic or inorganic material.
Many different microstructures have been exploited in an attempt to access the
intrinsic optical nonlinearity of metals, for example, in random
metallodielectric composites,\cite{shalaev02, shalaev96} fractal films,
\cite{shalaev96} and alternative bilayers,\cite{sipe95} etc. They basically
rely on the enhanced local fields in space or on the effectively lengthened
scale of the interactions between the matter and the incident light field.
However, there is also a great demand for particular optical materials in
devices applications, which would benefit from additional tunability of the
optical properties. Recently, we studied graded composites, which provide an
extra degree of freedom for controlling the optical properties of these
materials. \cite{Dong03} In fact, there exist in nature abundant graded
materials, such as biological cells \cite{gradedcell} and liquid crystal
droplets. \cite{dropt} Furthermore, many artificially-graded-index optical
metamaterials and elements have been fabricated nowadays. \cite{Levy05}

Colloidal crystal has been extensively studied in nanomaterials engineering
and its potential applications range from nanophotonics to chemistry and
biomedicine.\cite{Caruso05} It is desirable and of interest to use
dielectric-coated metallic nanoparticles with varying shell thickness to form
a dielectric constant gradient.\cite{Chan05}
%we put forth some kind of colloidal crystalline
%materials containing graded nanoparticle (graded colloid) \cite{HuangAPLnew} and that having a spatial variation
%of the nanoparticles shell structure (spatially graded colloid).\cite{Xiao} The resultant optical properties in
%the presence of the gradient in the colloid appear fascinating.
In this Letter, we theoretically investigate a metallic colloidal crystal
immersed in a graded-index host and demonstrate a giant enhanced optical
nonlinearity band, which is controllable by the gradient and by the
easily-tunable colloidal structure as well. These colloidal crystals can be
prepared via layer-by-layer self-assembly, templated sedimentation, methods
based on capillary forces, and electrokinetics. \cite{Caruso05, Chan05,
Velikov02, Gong04,Lumsdon03} We basically use the quasistatic point-dipole
approximation, which suffices in terms of characterizing both the gradient
effects and the lattice effects, otherwise the solution is formidable, either
from a Green's function formalism or from first-principles.\cite{Fst05}

The theoretical calculations are deployed on a model tetragonal lattice with
uniaxial anisotropy \cite{Lo03} (see Fig.~\ref{fig:stru}). As examples, let us
focus on the body-centered-tetragonal (bct), the body-centered-cubic (bcc) and
the face-centered-cubic (fcc) lattices respectively. Among the three cases bcc
lattice has the lowest packing density while fcc lattice has the highest one.
\cite{Lo03} Extensions to other colloidal structures such as simple tetragonal
lattices are straightforward and similar results are expected. Taking
advantages of the normalized interlayer interaction tensor $\textbf{T}$ (i.e.,
${\bf T}_{ij}$ denotes the interaction strength between two in-plane dipole
arrays) given by the Lekner summation method, \cite{Leknerer} we solve the
following self-consistent equation for the layer dependent local field
$\textbf{E}_i$
\begin{equation}
 \textbf{E}_i=\frac{1}{a^3}\sum_{j=1}^N {\bf{T}}_{ij} \cdot
(\alpha_j \textbf{E}_j)+\textbf{E}_i^{(0)}, \label{eq:localfield}
\end{equation}
where $a$ is the lattice constant ($b$ and $c$ are the other two lattice
constants) as shown in Fig.~\ref{fig:stru} , $\alpha_j$ is the layer-dependent
linear bare polarizability, here $i$, $j$ label the crystal layer and $N$
denotes the total layer number. Note $\textbf{E}_i^{(0)}$ in the
self-consistent equation is not simply the externally applied electric field
$\textbf{E}_0$ due to the presence of gradient. However, it is the field
inside the graded host medium, which is thus determined by virtue of the
continuity of the normal component of the electric displacement $\textbf{D}$
in the longitudinal case (i.e., $\textbf{E}_0$ parallel to the uniaxial axis).
It is the $z$-axis as shown in Fig.~\ref{fig:stru} in our case. Nevertheless
in the transverse case (i.e., $\textbf{E}_0$ perpendicular to the uniaxial
axis), we exactly used the applied field $\textbf{E}_0$ because the boundary
condition now becomes the continuity of the tangential component of electric
field. We shall compare the effective linear and nonlinear optical responses
of colloidal crystals with the different lattice structures (e.g., bct, bcc,
and fcc), made of metallic nanoparticles of linear dielectric constant
$\epsilon_1$ and third-order nonlinear susceptibility $\chi_1$, suspended in a
host fluid of $\epsilon_m$ (see Fig.~\ref{fig:stru}). Both the longitudinal
(L) and transverse (T) results will be discussed. The self-consistent
equations over $i=1, 2 \cdots, N$ are then combined together to take into
account the lattice effect and are being able to be transformed into a matrix
form as $\textbf{E=TAE+E}^{(0)}$. More precisely, in the longitudinal and the
transverse cases, $\textbf{E}=\{E_i^{(L,T)}\}$ is simply N-dimensional vector
and $\textbf{A}$ is $N\times N$ diagonal matrix of the polarizability,
%$\alpha_i (i=1, \cdots N)$,
%More precisely, in the longitudinal and the transverse case, it simply reads
%\begin{widetext}
%\begin{eqnarray}
%\left[\begin{array}{c} E_1\\E_2\\ \vdots \\E_N
%\end{array}\right]=
%\frac{1}{a^3} \left[
%\begin{array}{c c c c c}
%T_0 & T_1 & T_2 & \cdots &T_{N-1}\\
%T_1 & T_0 & T_1 & \cdots &T_{N-1}\\
%\vdots & \vdots & \vdots & \ddots & \vdots \\
%T_{N-1} & T_{N-2} &...&T_1&T_0 \\
%\end{array}\right]\left[
%\begin{array} {c c c c c}
%\alpha_1 & 0 & 0& \cdots & 0 \\
% 0 &\alpha_2 & 0& \cdots & 0 \\
%\vdots & \vdots & \vdots & \ddots & \vdots\\
%0  & 0& \cdots & 0&\alpha_N
%\end{array}\right]\left[
%\begin{array}{c}
%E_1\\E_2\\ \vdots \\E_N
%\end{array}\right]+
%\left[
%\begin{array}{c}
%E_1^{(0)}\\E_2^{(0)}\\ \vdots \\E_N^{(0)}
%\end{array}\right]
%\label{eq:locfldMatrix}.
%\end{eqnarray}
%\end{widetext}
%The agenda in this Letter mainly depends on Eq.~(\ref{eq:locfldMatrix}). In detail,
which relates the induced dipole moment $\textbf{p}_i$ of the particle and the
local field $\textbf{E}_i$ in the layer $\ell_i$, and in fact consists of
isotropic linear and nonlinear contributions. That is
%\begin{equation}
$\textbf{p}_i=\alpha_i\textbf{E}_i+\beta_i |\textbf{E}_i|^2\textbf{E}_i/3,
\label{eq:dipole}$
%\end{equation}
where
$\alpha_i=\epsilon_mr^3(\epsilon_1-\epsilon_m)/(\epsilon_1+2\epsilon_m)$. Here
$r$ is the radii of the metallic nanoparticles. In the case of weak
nonlinearity in the colloidal particles, i.e., $\chi_1|\textbf{E}_i|^2 \ll
\epsilon_1$ in the nonlinear relationship
$\textbf{D}_i=\epsilon_1\textbf{E}_i+\chi_1|\textbf{E}_i|^2\textbf{E}_i$, by
the perturbation expansion method,\cite{Yu92} one has
\begin{equation}
\beta_i=\left(\frac{3\epsilon_m}{\epsilon_1+2\epsilon_m}\right)^2\left
|\frac{3\epsilon_m}{\epsilon_1+2\epsilon_m}\right |^2r^3\chi_1.
\end{equation}
It is noteworthy that the linear local fields $\textbf{E}_i$ around the
particles in the layer $\ell_i$ are actually obtained by assuming no intrinsic
nonlinear response, i.e., we set $\chi_1=0$ for solving the self-consistent
equations [Eq.~\eqref{eq:localfield}], which is appropriate provided that the
nonlinear responses are much less than the linear ones. Next we use the
resultant linear local fields $\textbf{E}_i$ to extract the enhancement factor
of the effective nonlinear susceptibility $\overline{\chi}_1$
\cite{Yu92,Zeng88}
\begin{equation}
\gamma\equiv\frac{\overline{\chi}_1}{\chi_1}=
p\frac{\langle|E_i|^2E_i^2\beta_i\rangle}{3{|\bf{E}}_0^4|\chi_1},
\label{eq:enhancefactor}
\end{equation}
where $p$ is the filling fraction of the metallic colloidal nanoparticles. If
the nanoparticles are touching, i.e., with the geometric constraint
$a^2+b^2+c^2=16r^2$ (see Fig.~\ref{fig:stru}), then
$p=\pi[(q^3+2)/q]^{3/2}/24$ where $q=(c/a)^{2/3}$ quantifies the degree of
anisotropy of the periodic lattice \cite{Lo03} and also determines the
normalized interlayer interaction $\textbf{T}$. This results in
structure-controllable optical properties. Note that the average
$\langle...\rangle$ in Eq.~(\ref{eq:enhancefactor}) is taken over the layers
$\ell_i$ ($i=1,\cdots, N$) instead of over the nanoparticles spatial volume,
because in the dipole approximation the local fields inside each of the
particles are homogeneous. We also assume no nonlinear response in the host,
which is in fact relatively neglectable comparing to that in the metal.
Additionally, a gradient of the dielectric constant of the host fluid is
introduced along the uniaxial direction of the colloidal crystal, i.e,
$\epsilon_m=\epsilon_m(z_i)$ in our case, where $z_i$ represents the
$z$-coordinate of layer $\ell_i$. In this regard, we treat the host as a
continuously-layered film, thus explicitly has
$\textbf{E}^{(0)}\equiv\{E_i^{(0)}=E_0/\epsilon_m(z_i)\}$ in the longitudinal
case. The formation of the gradient in the host might be achieved by
dispersing different polymers in it, by selectively filling with microfluidic
materials,\cite{Sharkawy05} or induced by the presence of a temperature
gradient, etc. One can also simply coat the nanoparticles with different
coverage shells. But it still remains a challenge because the novel properties
from our prediction require a reasonably large gradient in the dielectric
constant of the host.

Figure~\ref{fig:fig2} shows in logarithmic scale the optical absorption, i.e.,
the imaginary part of the effective linear dielectric function
$\text{Im(}\epsilon_{\text{eff}}\text{)}$ (upper panels) and the modulus of
the nonlinearity enhancement factor $\gamma$ (lower panels) defined in
Eq.~(\ref{eq:enhancefactor}), as functions of the reduced frequency. We
specifically compare the results of bct ($q=0.87358$), bcc ($q=1.0$), and fcc
($q=1.25992$) as shown in the three columns for $N=25$ layers. The dielectric
function of the metallic colloidal nanoparticles is simply adopted as the
Drude form $\epsilon_1=1-\omega_p^2/(\omega^2+i\omega\Gamma)$, while the
graded dielectric constant in the host medium is assumed as $\epsilon_m(z_i) =
1.0 + 1.25i/N$ for $i=1, 2, \cdots N$. The presence of the inhomogeneity in
the host fluid obviously leads to a broadened and giant enhanced resonant band
(solid lines) in the low-frequency region. This is interesting for potential
telecommunication applications. The results of the same colloid crytal
suspended in homogeneous host medium with $\epsilon_m=1.0$ (dotted lines) or
$\epsilon_m=2.25$ (dashed lines) are also presented, in an attempt to
demonstrate that the broadened resonant band in some sense stems from the
hybridization of the non-graded structures. From the absorption spectrum and
the enhancement in the third-order nonlinear susceptibility, we would expect
an attractive figure of merit.\cite{Dong03} That is, the materials effectively
exhibit large nonlinearity and relatively small absorption. This is certainly
superior to pure metal because it generally has large nonlinearity and
unwanted absorption concomitantly. In fact, the optical absorption arises from
the surface plasmon resonance, which is obtained from the imaginary part of
the effective linear dielectric constant $\epsilon_{\text{eff}}$ that is
extracted from the generalized Clausius-Mossotti formula. \cite{Fst05, Lo03}
Note that the plasmon resonant peaks in the cases of homogeneous host with
periodic arrangment of metallic particles, i.e., the peaks in the dashed and
dotted curves in Fig.~\ref{fig:fig2} are red-shifted with respect to the
corresponding ones (not shown) of a homogeneous host with randomly-dispersed
nanoparticles predicted by the Maxwell-Garnett theory. We actually set a gap
(e.g., coated layer thickness in experiments) of $2a/5$ between the nearest
in-plane lattice particles in order to avoid severe complications arising from
the multiple image interactions. \cite{Fst05,Xiao05} Thus we have the metal
filling fraction $p=(3/5)^3 \pi[(q^3+2)/q]^{3/2}/24$, e.g., $0.15080$,
$0.14692$, and $0.15994$ for the bct, bcc, and fcc lattices, respectively. The
introduction of this gap indeed makes the nanoparticles size, and thus the
dipole factor relatively small and somewhat unfavorably suppresses the effect
arising from variation of the lattice structure, as seen in
Fig.\ref{fig:fig2}.

Furthermore, due to the fact that we treat the continuous variations of
dielectric function in the host as layered ones in obtaining $\alpha_i$ and
$\beta_i$, and the fact that the dipoles are actually distributed in discrete
lattice nodes, a series of sharp peaks are also observable in
Fig.~\ref{fig:fig2}. The peaks are merged in the broadened band and become
more notable for increased gradient in the host dielectric constant (not
shown), whereas they tend to disappear as the number of crystal layers $N$ is
increased. This is also understandable in the generalized Bergman-Milton
spectral representation in graded composites. \cite{Dong05} In detail, the
merging of the peaks into a continuous band is explicitly shown in
Fig.~\ref{fig:tune}, where we increased the layer to $N=50$. The fcc lattice
is taken as an example and we present both the longitudinal (left panel) and
the transverse (right panel) results. The peaks in Fig.~\ref{fig:fig2}(c) and
Fig.~\ref{fig:fig2}(f) are distinctly smeared out in Fig.~\ref{fig:tune}(a)
and Fig.~\ref{fig:tune}(b), respectively. The transverse results [see
Fig.~\ref{fig:tune}(c) and (d)] in the presence of the gradient is slightly
different to that in the longitudinal case [Fig.~\ref{fig:tune}(a) and (b)],
but still retain the broadened bands. We ascribe this to the fact that the
layer-to-layer interactions fall off exponentially due to the screening effect
in the lattice,\cite{Leknerer} therefore give no much
layer-structure-dependent difference in the two cases. Note that the
longitudinal and transverse results of crystals in homogeneous host (dotted
and dashed lines in Fig.~\ref{fig:tune}) do not differ much as well.

In conclusion, we theoretically exploit the optical resonant enhancement due
to lattice effect and gradient effect in colloidal crystals, which are made
out of suspended metallic nanoparticles in a graded-index host. The gradient
in the fluid and the colloid structure are easily subjected to tunability, for
example, the structure transformation might be induced by electrorheological
effects or by self-assembly of two kinds of particles with biochemically
different surface properties, etc. In addition, one can also use metal-covered
mangetic nanoparticles and control the suspension structure by external
magnetic field, consequently realizing magneto-controlled optical
properties.\cite{HuangAPL05} In this case, the electro-magnetorheological
effects is a good candidate as well. Devices that could benefit from these
materials include optical switches, optical limiters, as well as biosensors,
etc.

\hfill

This work was supported by the RGC Earmarked Grant. We thank Dr. J. P. Huang
for critical reading of the manuscript.

\newpage
\begin{center}
\textbf{Figure Captions}
\end{center}

\begin{figure}[htbp]
\caption{Schematic diagram of the colloidal crystal immersed in a graded host medium.
In the numerical calculations $a=b$ is assumed, which forms a square dipole
lattice in $xy$-plane, while $c$ is the lattice constant in the direction of
$z$-axis.} \label{fig:stru}
\end{figure}

\hfill

\begin{figure}[htbp]
\caption{The effective linear response $\text{Im(}\epsilon_{\text{eff}}\text{)}$
and third-order nonlinearity enhancement factor $|\gamma|$ of the periodic
colloidal nanoparticles immersed in a graded host fluid (solid line), and in
homogeneous host fluids with dielectric constant $\epsilon_2=2.25$ (dashed
line) and $\epsilon_2=1.0$ (dotted line), respectively; (a), (d) for bct, (b),
(e) for bcc; and (c), (f) for fcc, respectively. Parameters:
$\Gamma=0.02\omega_p$, $N=25$, $a=1$, and $|\textbf{E}_0|=1$. }
\label{fig:fig2}
\end{figure}

\hfill

\begin{figure}[htbp]
\caption{Same as Fig.~\ref{fig:fig2}, but with totally $N=50$ layers for fcc lattice.
(a) and (b) the longitudinal case; (c) and (d) the transverse case.}
\label{fig:tune}
\end{figure}

\newpage
\centerline{\includegraphics[scale=0.7]{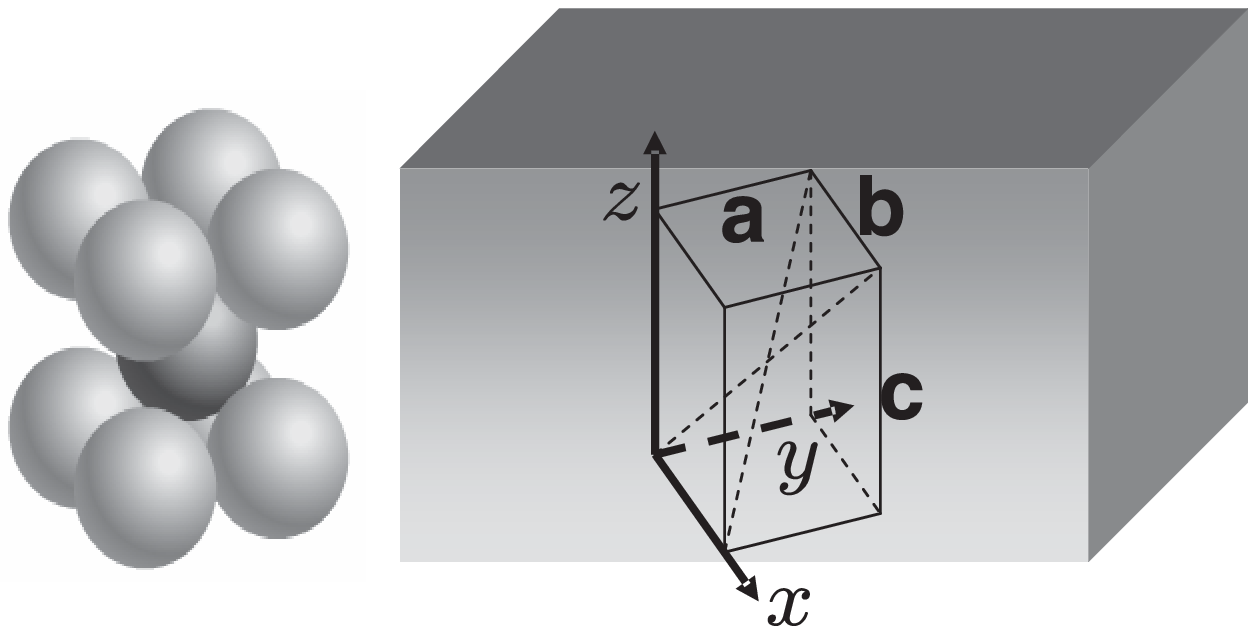}}
\centerline{Fig.1./Xiao and Yu}

\newpage
\centerline{\includegraphics[scale=0.8]{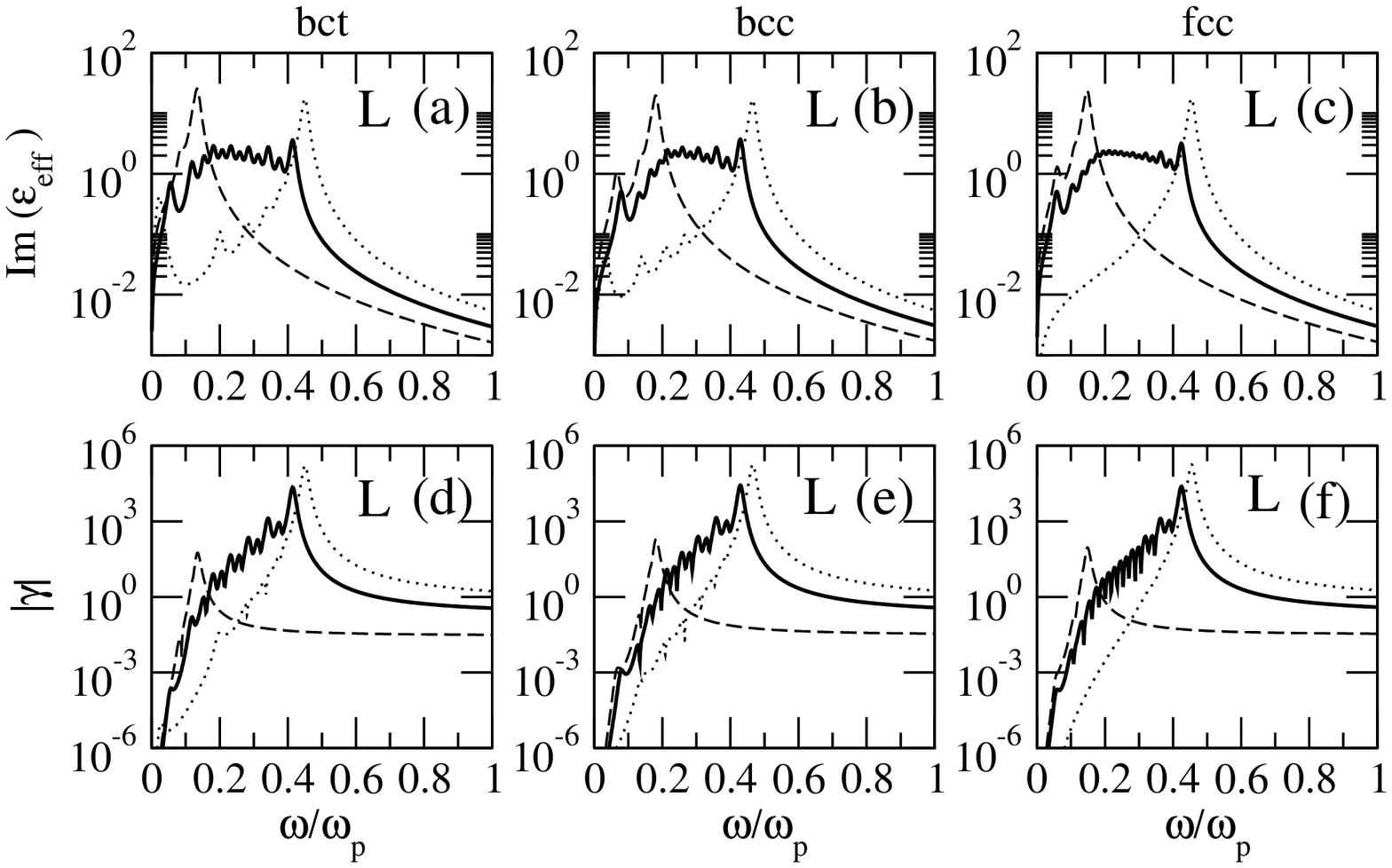}}
\centerline{Fig.2./Xiao and Yu}

\newpage
\centerline{\includegraphics[scale=0.7]{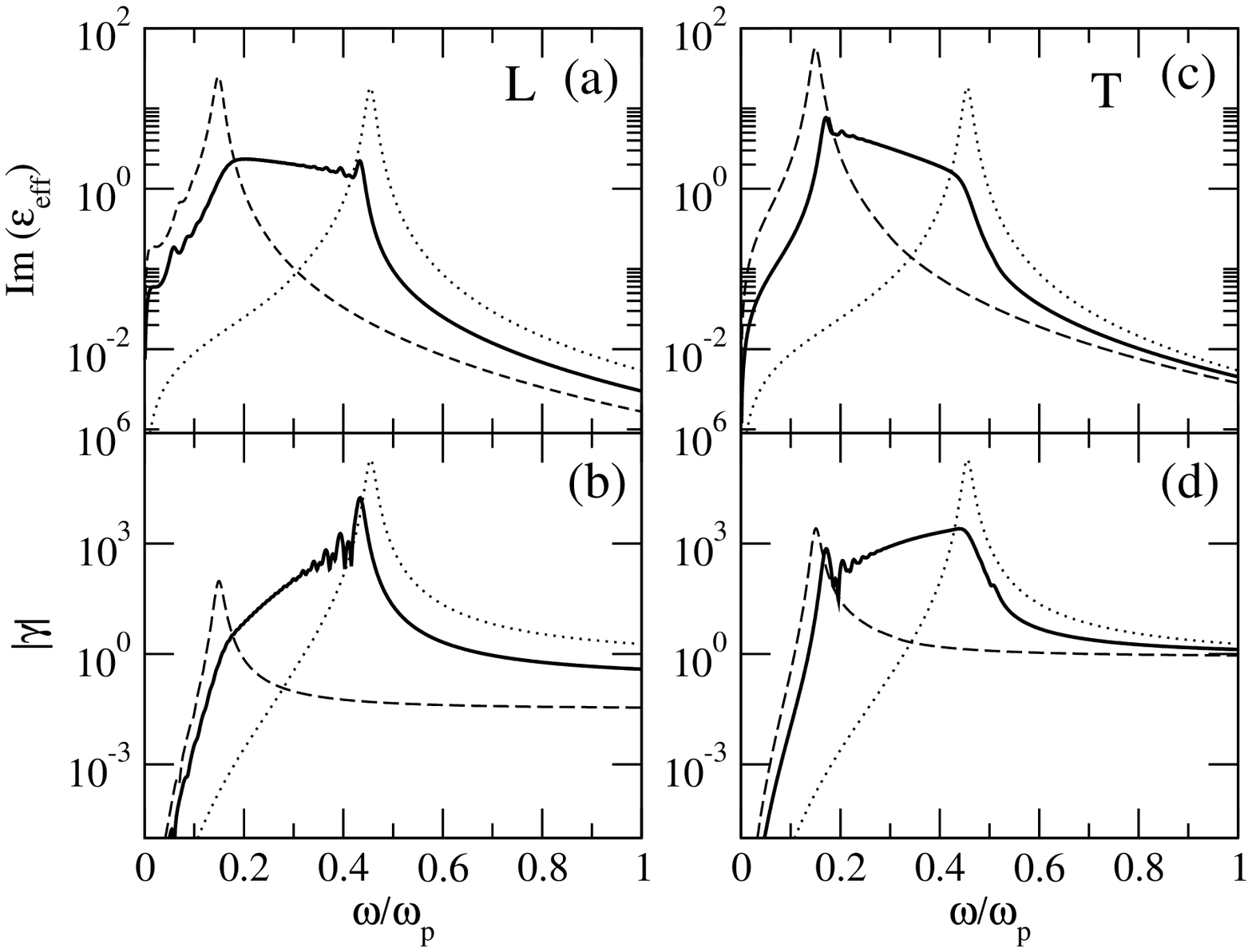}}
\centerline{Fig.3./Xiao and Yu}
\end{document}